\documentclass[a4paper,twoside]{article}

\usepackage{hyperref}
\usepackage{url} 
\usepackage{epsfig}
\usepackage{subcaption}
\usepackage{calc}
\usepackage{amssymb}
\usepackage{amstext}
\usepackage{amsmath}
\usepackage{amsthm}
\usepackage{multicol}
\usepackage{pslatex}
\usepackage{apalike}
\usepackage{algorithm2e}
\usepackage[bottom]{footmisc}
\usepackage{SCITEPRESS}     

\begin{document}

\title{Why are we living the age of AI applications right now?\\The long innovation path from AI's birth to a child's bedtime magic}

\author{\authorname{Tapio Pitkaranta\sup{1} }
\affiliation{\sup{1}Department of Computer Science and Engineering,\\ Aalto University, Finland, URL: http://www.aalto.fi }
\email{tapio.pitkaranta@iki.fi}
\href{https://www.linkedin.com/in/tapio-pitkaranta/}{LinkedIn: https://www.linkedin.com/in/tapio-pitkaranta/}
}


\abstract{Today a four-year-old child who does not know how to read or write can now create bedtime stories with graphical illustrations and narrated audio, using AI tools that seamlessly transform speech into text, generate visuals, and convert text back into speech in a natural and engaging manner. This remarkable example demonstrates why we are living in the age of AI applications. This paper examines contemporary leading AI applications and traces their historical development, highlighting the major advancements that have enabled their realization. Five key factors are identified: 1) The evolution of computational hardware (CPUs and GPUs), enabling the training of complex AI models 2) The vast digital archives provided by the World Wide Web, which serve as a foundational data resource for AI systems 3) The ubiquity of mobile computing, with smartphones acting as powerful, accessible small computers in the hands of billions 4) The rise of industrial-scale cloud infrastructures, offering elastic computational power for AI training and deployment 5) Breakthroughs in AI research, including neural networks, backpropagation, and the "Attention is All You Need" framework, which underpin modern AI capabilities. These innovations have elevated AI from solving narrow tasks to enabling applications like ChatGPT that are adaptable for numerous use cases, redefining human-computer interaction. By situating these developments within a historical context, the paper highlights the critical milestones that have made AI's current capabilities both possible and widely accessible, offering profound implications for society.}

\onecolumn \maketitle \normalsize \setcounter{footnote}{0} \vfill

\section{\uppercase{Introduction}}
\label{sec:introduction}

\begin{figure}[!h]
    \centering
\includegraphics[width=0.5\textwidth]{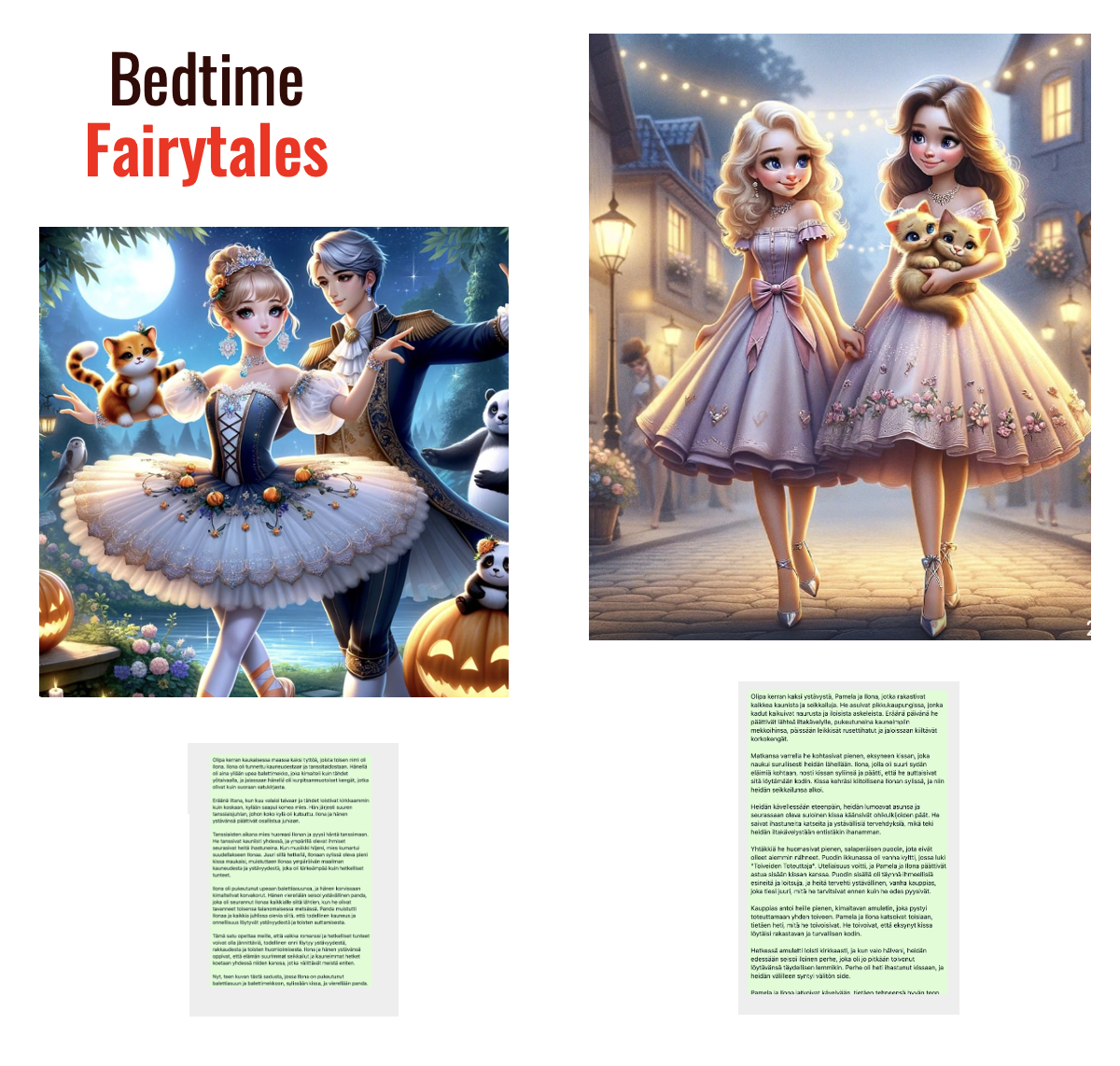}
	\caption{Bedtime stories created by a four year old child without adult help. The child does not know how to read or write. Stories are about one page long in total.}
\label{fig:bedtime-stories}
\end{figure}

\begin{figure}[!h]
    \centering
\includegraphics[width=0.5\textwidth]{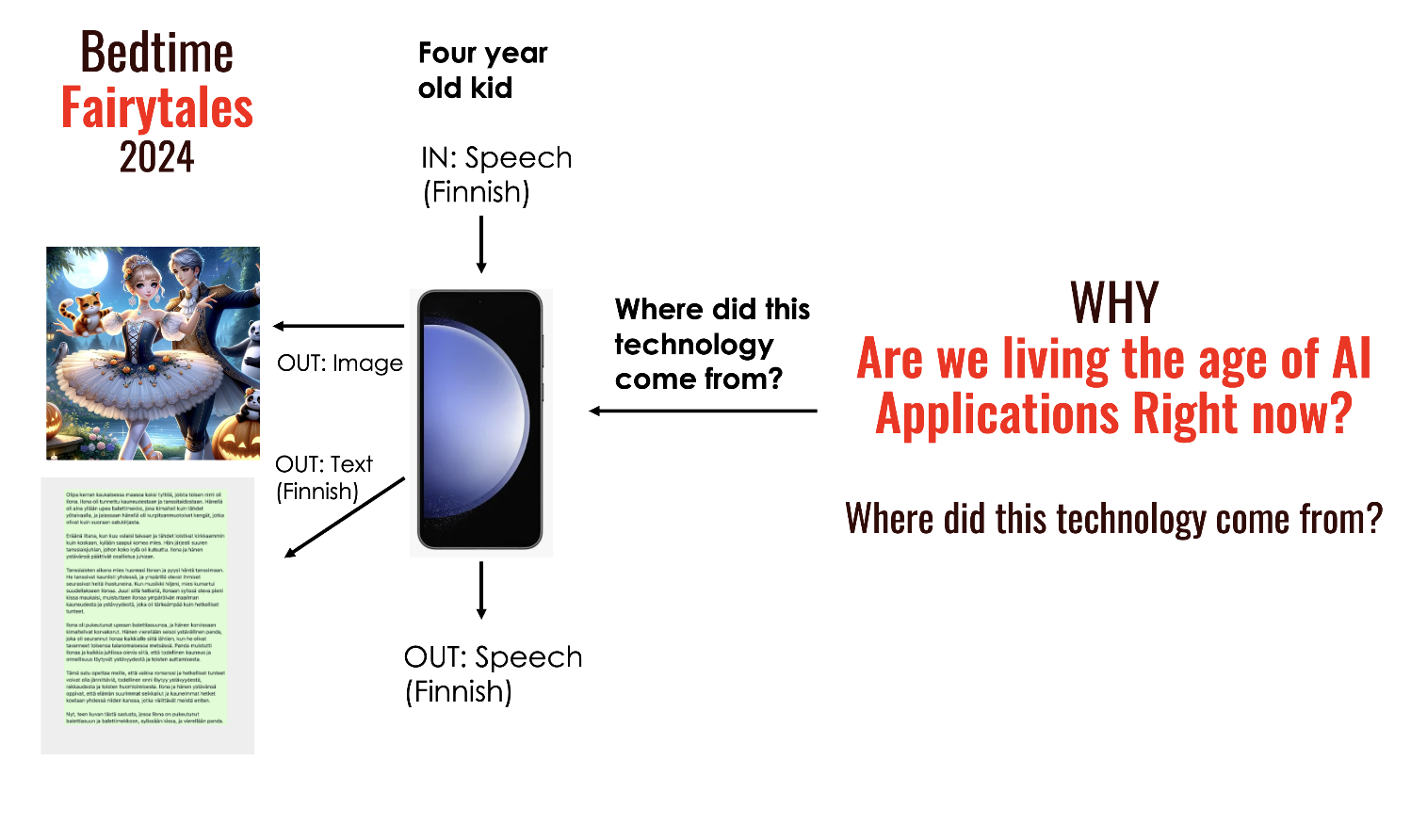}
	\caption{Research question in this paper: Where did this technology come from?}
\label{fig:where-did-openai-android-come-from}
\end{figure}

For a little over a year, a four-year-old child has been creating bedtime stories, complete with illustrations, simply by speaking into a phone. Real examples of these stories can be seen in Figure~\ref{fig:bedtime-stories}. These are not just any stories; the four-year-old, who does not yet know how to read or write, has a remarkably detailed vision of what the stories should contain. After generating the visuals and listening to the story spoken by the AI, the AI often needs to adjust the pictures or the narrative because the initial result does not precisely match the child's vision. The adult is not technically needed in this interaction; however, I have chosen to remain present to ensure that both the AI and the child are engaging appropriately.

This remarkable example demonstrates why we are living in the age of AI applications. This remarkable capability - enabled by advanced artificial intelligence (AI) - highlights the profound transformation in human-computer interaction and the accessibility of complex technologies to even the youngest members of society. Such seamless integration of AI into everyday life demonstrates the culmination of decades of scientific breakthroughs and technological advancements.

This paper examines the convergence of technological and scientific advancements that have enabled such transformative experiences. We begin with this state-of-the-art AI application used by the four-year-old child, which actually comprises multiple small AI features. In this paper, we aim to trace backwards through history to identify the main scientific breakthroughs that made this AI application possible. The approach of the paper is illustrated in Figure~\ref{fig:where-did-openai-android-come-from}.

This paper seeks to explore the historical and technological underpinnings that have made this extraordinary scenario possible. By examining the trajectory of key developments in AI research, computational infrastructure, data accessibility, and communication technologies, we aim to trace the critical milestones that have collectively shaped the modern AI landscape. The purpose of this retrospective analysis is to identify the foundational achievements and pivotal moments that brought AI to its current state of ubiquity and general utility.

Through this exploration, we provide a framework for understanding how the convergence of scientific discovery and technological progress has led to the democratization of AI, where even the most complex applications are accessible to a child. This context not only underscores the significance of AI's evolution but also sets the stage for appreciating its broader implications for society and future innovation.

\subsection{Research Questions}
\label{sec:research_questions}

\begin{enumerate} 

\item Why are we living the age of AI right now?

\item What are the major hardware and software components of this technology? 

\item What are the major AI capabilities of this technology? 

\item What historical technological advancements have contributed to the development of this technology? 

\item What are the major scientific breakthroughs that have enabled this technology?

\end{enumerate}

\subsection{Research Objectives}
\label{sec:research_objectives}

\begin{enumerate} 

\item Systematically address the research questions to provide a comprehensive understanding of the technology.

\item Identify and analyze the minimum essential set of scientific breakthroughs and advancements critical to the development of this technology, focusing on quality over quantity.

\item Trace the historical evolution of the technology, uncovering patterns and interdependencies that highlight how key components and ideas emerged.

\item Utilize academic and web-based resources, supported by generative AI tools, to iteratively refine insights and improve the quality of the research process.

\end{enumerate}

\section{\uppercase{Research Method}}

The research method employed in this paper begins with the use of OpenAI software running on an Android device as depicted Figure~\ref{fig:where-did-openai-android-come-from}. The approach involves iteratively addressing research questions by systematically analyzing the main components of the technology, progressing layer by layer to uncover original ideas or scientific breakthroughs underpinning its development.

Once the main components are identified, academic search engines such as Google Scholar are utilized to locate original research papers and foundational ideas. Subsequently, web searches are conducted to trace the historical evolution of the relevant technologies, identifying patterns in their emergence and development.

Generative AI served as a chat assistant throughout the research process, facilitating the iteration of ideas and ensuring grammatical correctness in the text. This iterative approach integrates structured academic research with advanced AI tools to enhance the depth and accuracy of the findings. Speech-to-text AI has been employed to convert the initial versions of the paper from spoken to written format. A large language model has been utilized to summarize key points from the transcripts for the purposes of this paper.

\section{\uppercase{Why are we living the age of AI right now}}

\subsection{Brake down of technology components}

\begin{figure}[!h]
    \centering
\includegraphics[width=0.5\textwidth]{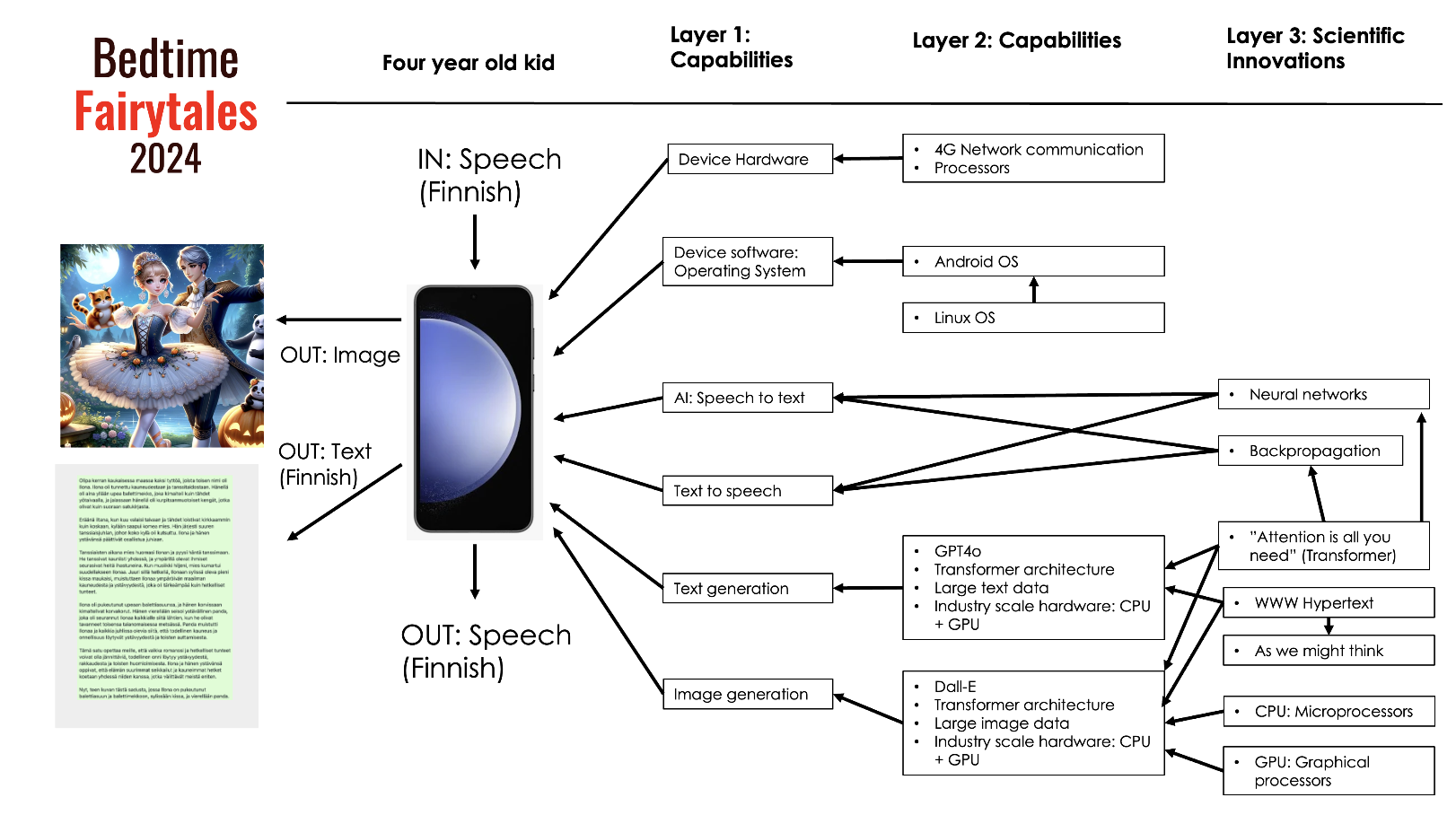}
	\caption{Brake down of main technology components that are needed for the AI application used for bedtime story generation}
\label{fig:where-did-openai-android-come-from-main-components}
\end{figure}

\subsection{Segmentation of the technologies}

Five key factors are identified: 

\begin{enumerate} 
\item \textbf{The evolution of computational hardware (CPUs and GPUs),} enabling the training of complex AI models 

\item \textbf{The vast digital archives provided by the World Wide Web,} which serve as a foundational data resource for AI systems

\item \textbf{The ubiquity of mobile computing,} with smartphones acting as powerful, accessible devices in the hands of billions

\item \textbf{The rise of industrial-scale cloud infrastructures,} offering elastic computational power for AI training and deployment 

\item \textbf{Breakthroughs in AI research,} including neural networks, backpropagation, and the \textit{Attention is All You Need} framework, which underpin modern AI capabilities. 

\end{enumerate} 

These innovations have elevated AI from solving narrow tasks to enabling general-purpose applications like ChatGPT, redefining human-computer interaction. By situating these developments within a historical context, the paper highlights the critical milestones that have made AI's current capabilities both possible and widely accessible, offering profound implications for society.

\subsection{AI Calculations: Modern Processors (CPU) and Graphics Cards (GPU)}

\begin{figure}[!h]
    \centering
\includegraphics[width=0.5\textwidth]{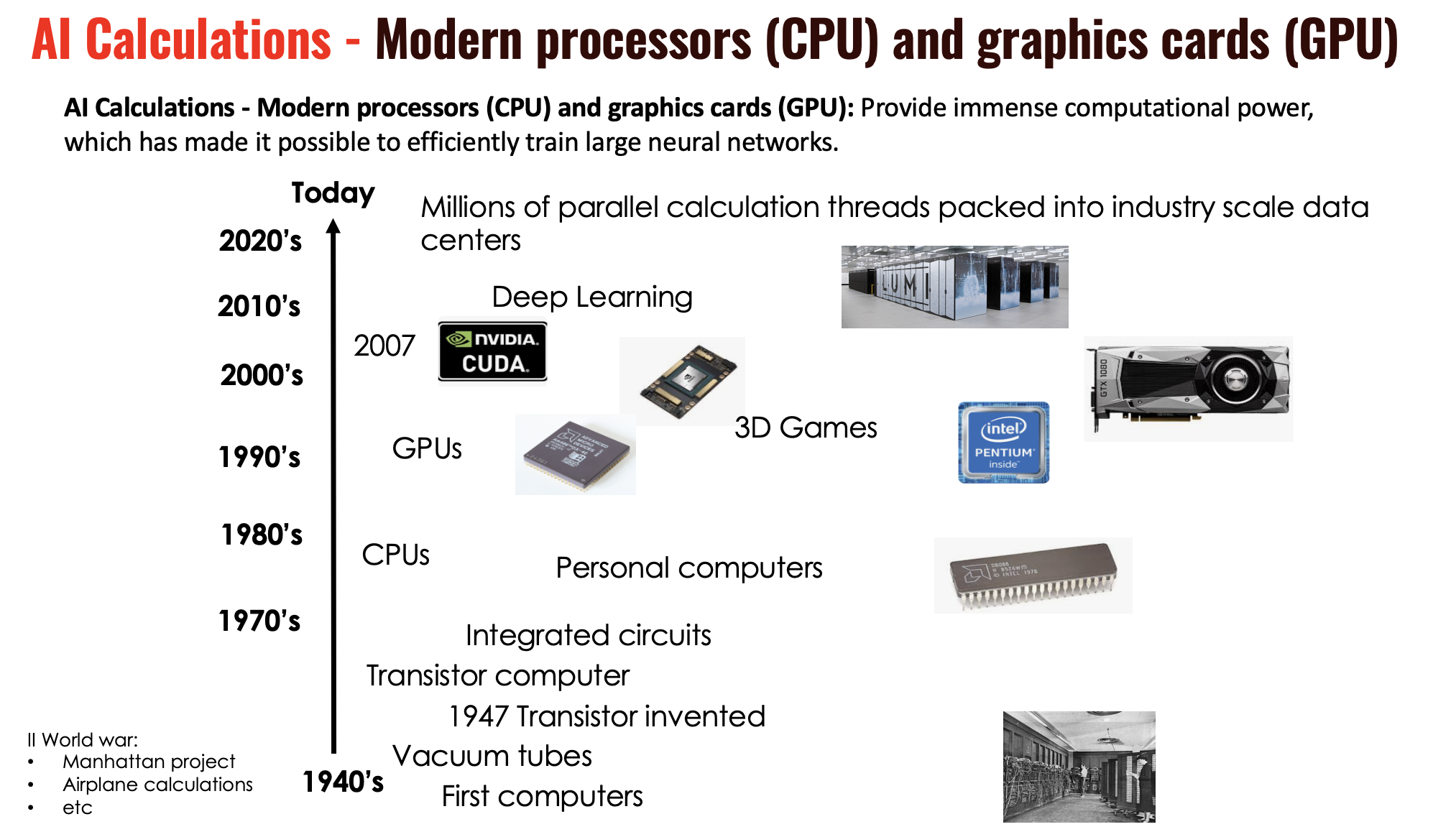}
	\caption{AI Calculations - Modern processors (CPU) and graphics cards (GPU): Provide immense computational power, which has made it possible to efficiently train large neural networks.}
\label{fig:cpu-gpu-history}
\end{figure}

The evolution of computational hardware has been instrumental in the development of artificial intelligence (AI) technologies. This subsection explores the role of modern processors, including central processing units (CPUs) and graphics processing units (GPUs), in advancing AI calculations.

\subsubsection{Early Processors and Their Evolution}

In the early stages of computing, processors were designed to handle basic numerical calculations. ENIAC (1945), one of the earliest programmable electronic computers, relied on vacuum tubes to perform a range of calculations, albeit requiring manual rewiring for different tasks. The IBM 704 (1954) introduced hardware support for floating-point numbers, making it a significant milestone in enabling complex mathematical computations.

The invention of microprocessors in the 1970s marked a turning point in computing. Intel's 4004 processor (1971) was the first commercial microprocessor, paving the way for personal computing. This was followed by the Intel 8086 (1978), which became the industry standard for PCs. These advancements laid the groundwork for more sophisticated computing devices, including AI-enabling technologies.

\subsubsection{The Rise of GPUs for AI}

The history of GPUs (Graphics Processing Units) reflects decades of innovation, beginning in the 1970s with the advent of specialized graphics hardware for rendering vector and raster graphics. These early developments laid the foundation for interactive computer graphics used in fields such as CAD and gaming. In the 1980s, advancements like framebuffer technology and the introduction of hardware-accelerated 2D and 3D graphics systems enabled more efficient rendering of complex scenes, with products such as IBM's Professional Graphics Controller (1984) marking a key milestone. The 1990s saw the emergence of fully integrated GPUs, with NVIDIA's GeForce 256 (1999) heralded as the first "GPU," introducing on-board transform and lighting capabilities and defining a new standard for graphics computation~\cite{wikipedia_gpu_history}. In the 2000s, GPUs evolved beyond graphics rendering, becoming indispensable for parallel computing tasks, including scientific simulations and machine learning, driven by the programmability introduced with NVIDIA's CUDA architecture (2006). This transformation positioned GPUs as critical components in fields ranging from gaming to artificial intelligence and high-performance computing.

\subsubsection{GPUs: From Graphics to General-Purpose Computing}

Initially developed for rendering images in video games, GPUs underwent a significant transformation with the introduction of programmability. The NVIDIA GeForce 256, introduced in 1999, marked a pivotal moment, as it demonstrated the capability of GPUs to handle parallel computation. This innovation laid the groundwork for general-purpose GPU computing (GPGPU), which gained momentum in the 2000s with the release of NVIDIA CUDA in 2007. CUDA provided developers with a flexible programming model to harness the inherent parallelism of GPUs for tasks beyond graphics, including scientific simulations, data processing, and machine learning.

\subsubsection{GPUs and the AI Revolution}

The evolution of GPUs into powerful tools for AI marked a critical shift in computing. Their ability to process multiple tasks concurrently has made them ideal for training large-scale neural networks, an essential component of modern AI. With the launch of products such as NVIDIA's A100 GPU in 2020, GPUs have been tailored specifically for AI and high-performance computing (HPC) workloads, offering unprecedented computational power and efficiency. This specialization has cemented GPUs as the backbone of AI development, driving advancements in fields ranging from natural language processing to autonomous systems.

\subsubsection{Specialized Accelerators and Quantum Computing}

The need for greater efficiency and computational power has led to the development of specialized accelerators, such as Google's Tensor Processing Unit (TPU). Introduced in 2016, TPUs are optimized for AI computations, particularly deep learning, and have significantly accelerated the training of large-scale models.

Quantum computing, though still in its nascent stages, represents the next frontier in computational hardware. By leveraging quantum mechanics, these machines promise exponential increases in processing power, which could revolutionize AI by enabling the solution of problems currently beyond the reach of classical computers.

\subsubsection{Scientific Breakthroughs Driving Hardware Utilization}

The synergy between hardware advancements and scientific breakthroughs has been pivotal in the AI revolution. Algorithms such as backpropagation, essential for training neural networks, and architectural innovations like the Transformer model ~\cite{vaswani2017attention} have capitalized on the capabilities of modern CPUs and GPUs. These innovations have enabled the development of large-scale models like ChatGPT, which rely on billions of parameters and vast computational resources.

In summary, the interplay between advanced computational hardware and AI algorithms has ushered in an era of unprecedented technological possibilities. From the foundational ENIAC to modern GPUs and TPUs, each generation of processors has contributed to the evolution of AI, laying the groundwork for future breakthroughs in computing and machine intelligence.

\subsection{Data for AI: WWW as the Killer Application for the Internet}

\begin{figure}[!h]
    \centering
\includegraphics[width=0.5\textwidth]{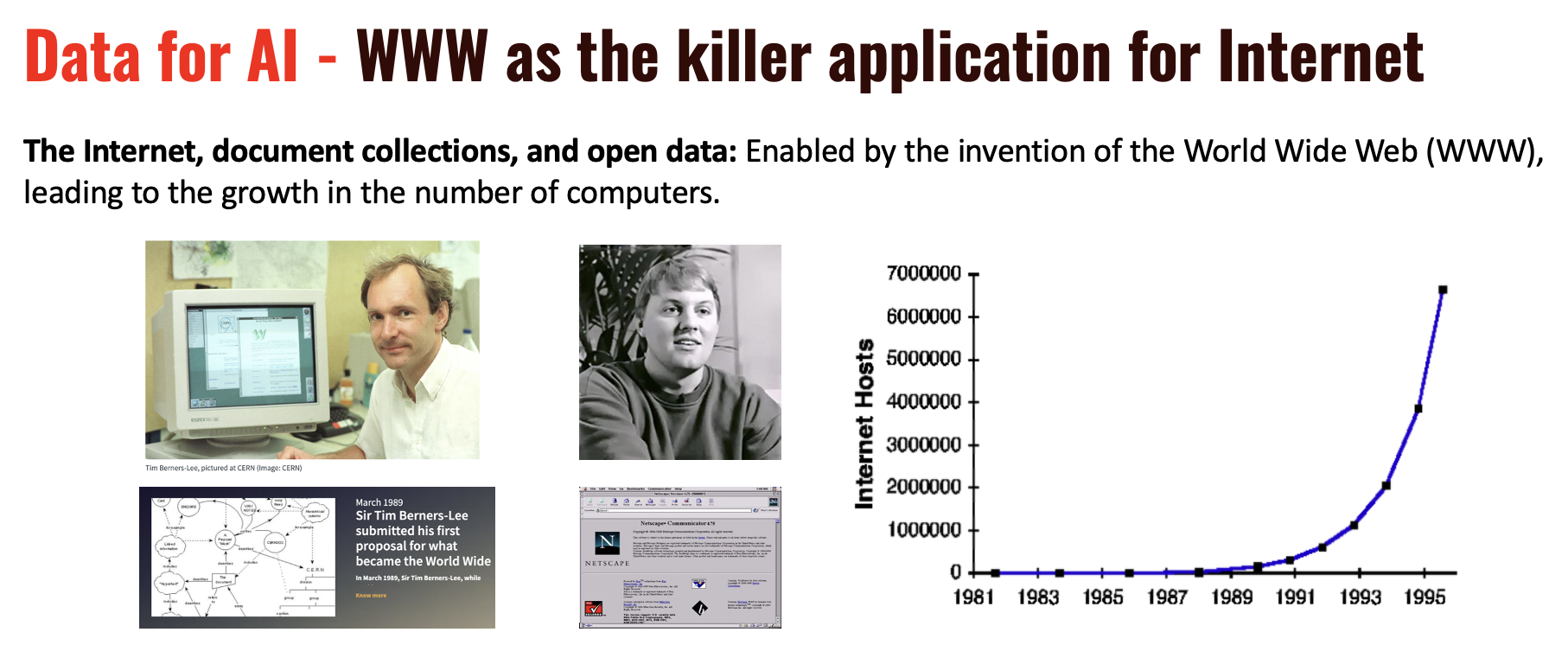}
	\caption{World wide web (WWW) as the killer application for the internet. Tim Berners-Lee invented the WWW and Marc Anreessen implemented one of the early commercial browser applications (Netscape)}
\label{fig:www-killer-app-for-internet}
\end{figure}

The emergence of the World Wide Web (WWW) in the early 1990s transformed the way data is created, accessed, and shared, creating the largest public interlinked (hypertext) data collection that humankind has ever had. Enabled by Tim Berners-Lee's vision in 1989~\cite{berners_lee_1989} and its public unveiling in 1991, the WWW revolutionized information sharing and laid the groundwork for the explosion of data essential for AI development.

Vannevar Bush is widely acknowledged to be the inventor of hypermedia. As early as in 1945 he wrote a famous article called \textit{As We May Think} in the Atlantic Monthly ~\cite{vanner45}, where he broadly discussed the problems of information management in the research world of those days. Douglas Engelbart's 1968 demonstration, often referred to as "The Mother of All Demos," showcased groundbreaking innovations such as the mouse, graphical user interfaces, hypertext, and collaborative computing \cite{engelbart:1968}.

\subsubsection{The Role of the WWW in Data Growth}

Between 1990 and 2000, the WWW dramatically increased the availability of document collections and open data, enabling rapid growth in computational applications. With the introduction of the Netscape Navigator in 1994, web access became more user-friendly, expanding the reach of the internet to a broader audience. This facilitated the collection of the largest repository of human knowledge, accessible globally, which became a cornerstone for AI training and research.

The WWW can be regarded as a "killer application" for the internet due to its transformative impact on communication, commerce, education, and collaboration. By enabling the accumulation and dissemination of structured and unstructured data, the web supported the development of numerous AI applications, such as natural language processing, recommendation systems, and computer vision.

\subsubsection{From Static Web to Dynamic AI Applications}

In the early stages of the WWW, static web pages provided a platform for information sharing. Over time, advancements in web technologies enabled interactive applications and data-driven services. The proliferation of user-generated content, such as social media posts, images, and videos, further enriched the data landscape, creating diverse datasets essential for training AI models.

Moreover, the advent of cloud computing in the late 2000s leveraged the WWW's infrastructure to enable scalable data storage and computation. Platforms such as AWS, Google Cloud, and Microsoft Azure offered the computational capacity to process the vast data generated via the web, fueling AI's growth and accessibility.

\subsubsection{Modern Search Engines Enabling WWW Data Consumption}

The World Wide Web (WWW) began to expand rapidly in the mid-1990s, becoming the largest interconnected collection of documents in human history. During this period, efficiently indexing, ranking, and retrieving relevant information from this vast and unstructured data posed a significant challenge.

The search engine market was transformed by the work of Brin and Page \cite{brin_page:1998}, who introduced the architecture of Google, featuring PageRank - a link analysis algorithm that leveraged the web's hyperlink structure to evaluate the importance of pages. This innovation allowed Google to deliver more relevant search results with improved speed and accuracy, marking a shift from simple keyword matching to understanding the structure of the web.

Google addressed the challenges of data retrieval while enabling the continued growth of the WWW by ensuring that a growing user base could efficiently find relevant information. By making web content more accessible, Google contributed to the expansion of the web and increased participation by individuals and organizations.

Google became a widely used interface for accessing information on the WWW, significantly shaping how people retrieve and use data. Its innovations in search laid the foundation for advancements in natural language processing, recommendation systems, and AI-driven search technologies.

\subsubsection{Implications for AI Development}

The success of the WWW catalyzed the exponential growth of data, a critical ingredient for AI advancements. Machine learning models require large, diverse datasets for training and validation, and the WWW provided an unparalleled source of such data. Furthermore, the open and collaborative nature of the web facilitated the sharing of research, algorithms, and tools, accelerating innovation in AI.

In summary, the World Wide Web was not only a transformative application of the internet but also an essential enabler of modern AI. By fostering unprecedented access to data and computational resources, it set the stage for the AI era, influencing technological, societal, and scientific advancements that continue to evolve.

\subsection{Calculation Scale: Development of Mobile and Communication Technology}

\begin{figure}[!h]
    \centering
\includegraphics[width=0.5\textwidth]{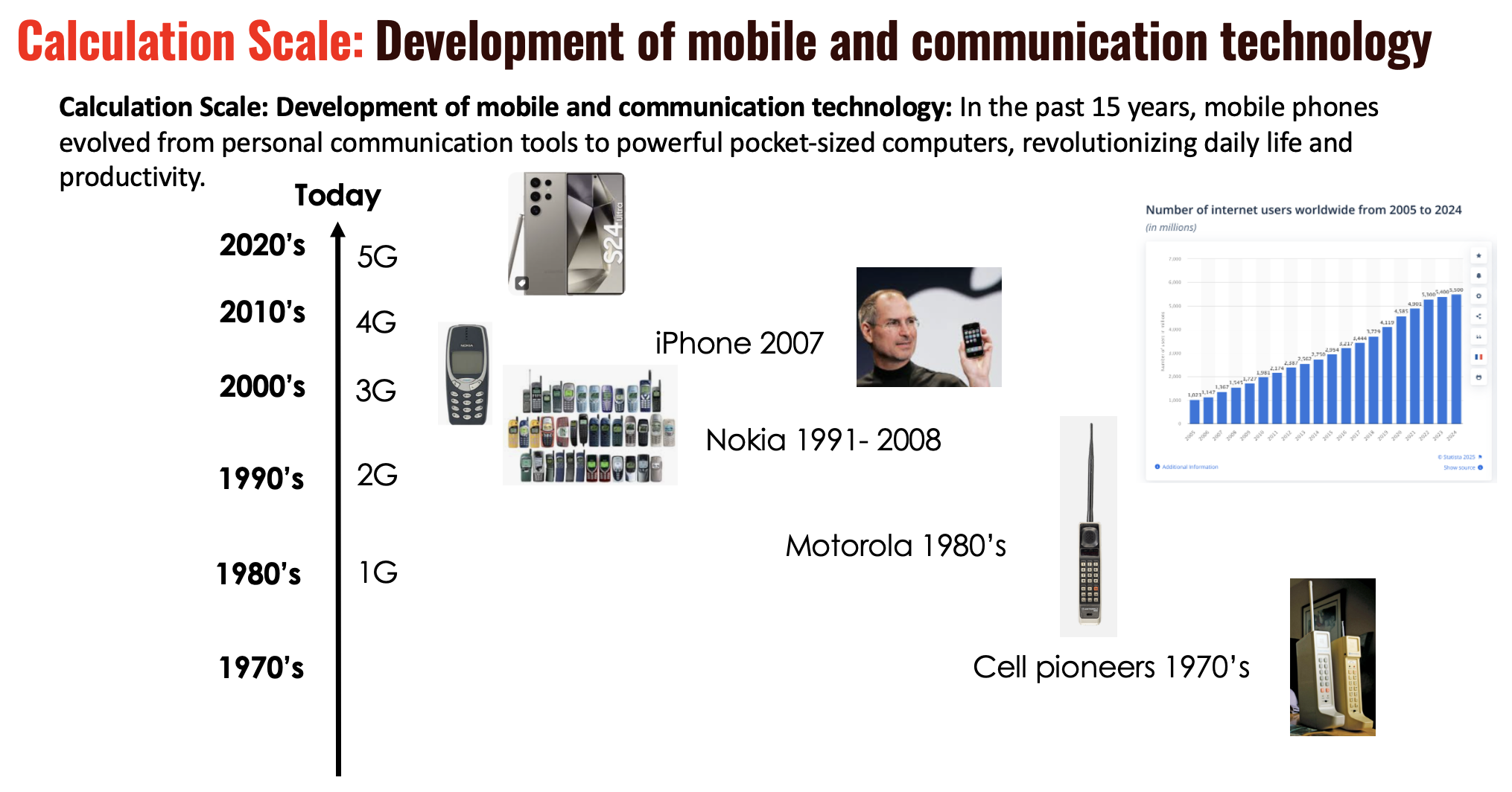}
	\caption{ Evolution of Mobile Phones: Today, the most advanced Android phones function as powerful computers in your pocket. The evolution of network technologies has progressed from 1G to the current 5G networks.}
\label{fig:mobile-phone-history}
\end{figure}

The evolution of mobile and communication technology has played a pivotal role in the modern computational era, transforming mobile phones from basic communication tools into powerful, pocket-sized computers. Over the past 15 years, these advancements have revolutionized daily life and productivity, driven by the development of successive generations of mobile networks and devices.

\subsubsection{Early Foundations (1900s - 1940s)}

The origins of mobile communication can be traced back to the early 20th century with the development of radio telephony for military and maritime use. In 1946, Bell Labs introduced the Mobile Telephone Service (MTS), a car-based system utilizing large radio towers, marking the first significant step toward mobile telephony. 

\subsubsection{The Birth of Cellular Networks (1947 - 1970s)}

In 1947, engineers at Bell Labs proposed the cellular concept, which divided service areas into "cells" with radio towers to enhance frequency reuse and improve capacity. Early car phones in the 1950s and 1960s, such as the Improved Mobile Telephone Service (IMTS), reduced equipment size but remained bulky and costly.

\subsubsection{The First Generation (1G) - Analog Systems (1980s)}

The 1980s marked the debut of handheld mobile phones with the launch of Motorola's DynaTAC 8000X in 1983. Despite its limited 30-minute talk time and high cost, it represented a major breakthrough in mobile technology. Concurrently, 1G networks like AMPS (Advanced Mobile Phone System) enabled voice-only calls, though with limited coverage and capacity.

\subsubsection{The Second Generation (2G) - Digital Revolution (1990s)}

The introduction of digital technology in the 1990s with 2G networks, such as GSM (Global System for Mobile Communications) and CDMA (Code Division Multiple Access), improved call quality and security. This period also witnessed the advent of text messaging, with the first SMS ("Merry Christmas") sent in 1992. Devices like the Nokia 1011 became smaller, more affordable, and widely adopted.

\subsubsection{The Third Generation (3G) - Mobile Internet (2000s)}

The early 2000s saw the proliferation of 3G networks, enabling faster data speeds and supporting email, video calls, and internet browsing. Smartphones began to emerge, with BlackBerry revolutionizing business communication, Nokia N-Series integrating cameras and multimedia capabilities, and Apple's iPhone (2007) introducing touchscreens and app ecosystems.

\subsubsection{The Fourth Generation (4G) - High-Speed Connectivity (2010s)}

The 2010s ushered in 4G networks, characterized by LTE (Long-Term Evolution), which facilitated high-speed streaming, gaming, and video calls. Smartphones became more powerful, featuring advanced cameras, high-resolution screens, and extensive app ecosystems. Operating systems like Google Android and Apple iOS dominated this era.

\subsubsection{The Fifth Generation (5G) - Ultra-Fast Networks (2020s)}

The 2020s have witnessed the advent of 5G networks, offering speeds up to 100 times faster than 4G with significantly reduced latency. These networks have enabled real-time applications that are competitive with many use cases traditionally reliant on cable or fiber connections. Furthermore, 5G supports billions of connected devices, driving innovation in smart homes, cities, and industries through the Internet of Things (IoT).

\subsubsection{Mobile Phone Operating Systems}

The Android operating system, a cornerstone of modern mobile computing, has evolved significantly since its launch in 2008 as an open-source platform \cite{wikipedia_android_history}. Designed to power a wide range of mobile devices, Android built upon the Linux kernel, which provided a stable and secure foundation for its architecture. The Linux kernel itself traces back to Linus Torvalds' groundbreaking 1991 announcement of Linux, a free and open-source Unix-like operating system \cite{wikipedia_history_linux}. Torvalds' contributions to open-source software laid the groundwork for subsequent innovations like Android, which leveraged Linux's robustness to create a versatile operating system for mobile devices. This lineage highlights the profound influence of Linux on the development of mobile operating systems and the broader computing landscape \cite{wikipedia_mobile_os}.

\subsubsection{Implications for AI and Beyond}

Today, there are over 6 billion smartphones - essentially small computers - in the world, all of which rely on vast background processing and data storage capabilities provided by hyperscale server farms, commonly known as cloud services.

The evolution of mobile technology has significantly influenced AI by enabling the collection and processing of vast amounts of data. Billions of smartphones worldwide act as both data generators and access points to cloud-based computational resources. As mobile and communication technology continue to advance, their integration with AI promises to drive further innovation across multiple domains.

\begin{figure}[!h]
    \centering
\includegraphics[width=0.5\textwidth]{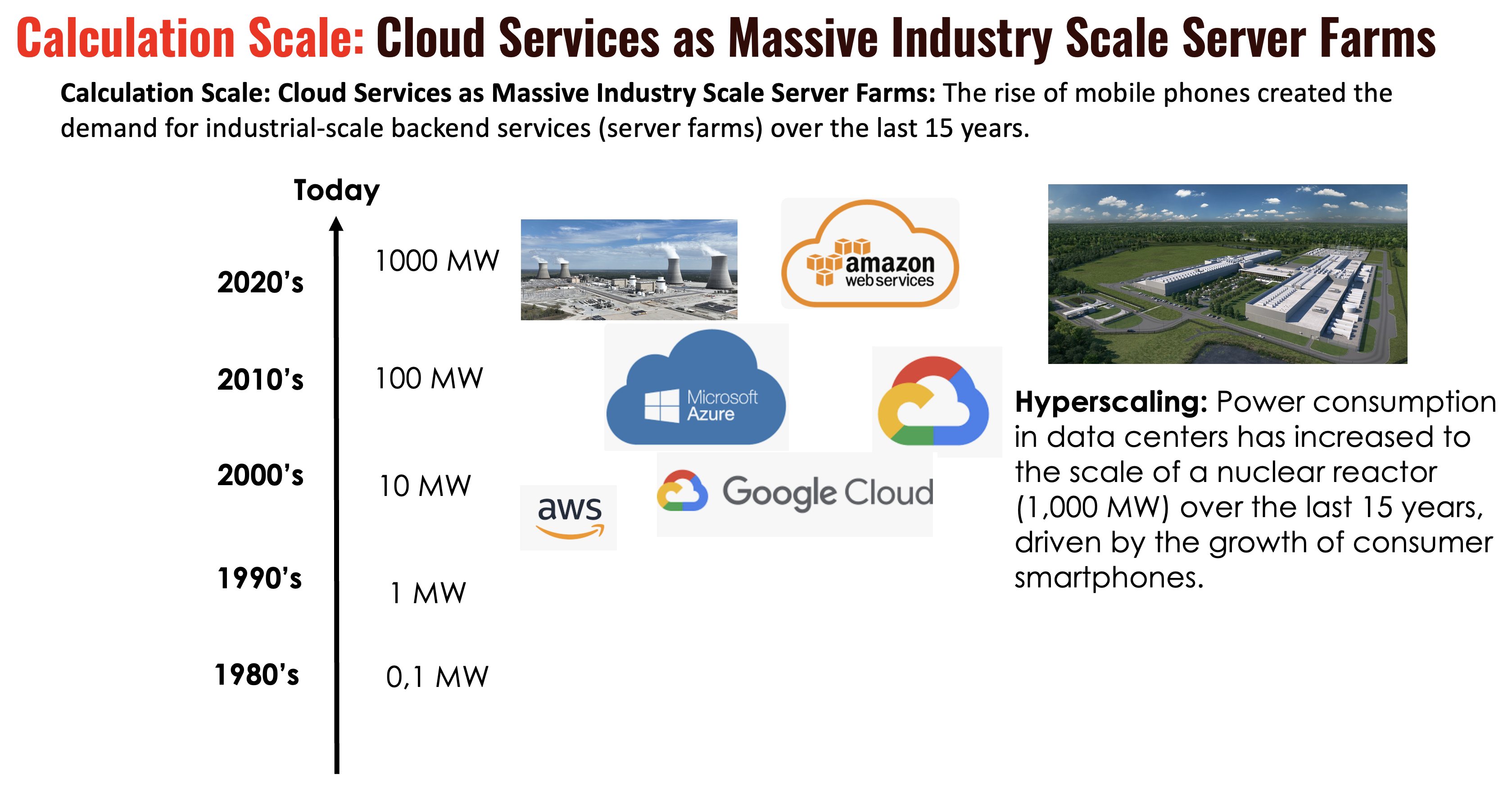}
	\caption{Hyperscaling: Power consumption in data centers has increased to the scale of a nuclear reactor (1,000 MW) over the last 15 years, driven by the growth of consumer smartphones.}
\label{fig:hyperscale-datacenters}
\end{figure}

\subsection{Calculation Scale: Cloud Services as Massive Industry Scale Server Farms}

The rise of mobile phones and the exponential growth of internet usage over the past two decades have necessitated the development of massive industrial-scale backend services, commonly referred to as cloud computing. These "server farms" provide scalable computational resources that have become the backbone of modern digital services, including AI and machine learning applications.

\subsubsection{Early 2000s: The Emergence of Cloud Computing}

The concept of cloud computing gained prominence in the early 2000s as businesses recognized the inefficiency of maintaining underutilized data centers. In 2006, Amazon Web Services (AWS) revolutionized the industry with the launch of Elastic Compute Cloud (EC2) and Simple Storage Service (S3), introducing Infrastructure as a Service (IaaS). AWS pioneered the hyperscaling model, which allowed businesses to pay only for the resources they used, transforming resource allocation and scalability.

\subsubsection{Late 2000s: Competitors Enter the Market}

The late 2000s witnessed the entry of major competitors into the cloud computing arena. In 2008, Google launched App Engine, focusing on Platform as a Service (PaaS) and later expanding to IaaS and hybrid solutions through Google Cloud Platform (GCP). Microsoft followed in 2010 with Azure, initially catering to PaaS developers before evolving into a comprehensive IaaS provider.

\subsubsection{2010s: Rapid Expansion and Hyperscaling}

During the 2010s, cloud platforms underwent rapid expansion, leveraging commodity hardware, virtualization, and containerization technologies like Docker and Kubernetes. Investments in global data center regions and content delivery networks (CDNs) enabled low-latency services worldwide. AWS solidified its dominance by continuously adding services for machine learning, databases, and analytics. The adoption of multi-cloud strategies further diversified the market, allowing businesses to leverage redundancy and unique features from multiple providers.

\subsubsection{Late 2010s: Edge Computing and Serverless Technologies}

The late 2010s marked a shift toward edge computing, with hyperscalers bringing computation closer to users for real-time analytics and Internet of Things (IoT) applications. Serverless computing models, such as AWS Lambda, allowed developers to execute code without managing the underlying server infrastructure, enhancing scalability and cost-efficiency.

\subsubsection{2020s: Dominance and Innovation}

The COVID-19 pandemic accelerated cloud adoption, fueling demand for remote work solutions, digital transformation, and online services. Hyperscalers invested heavily in advanced technologies like artificial intelligence (AI), machine learning (ML), quantum computing, and high-performance computing (HPC). The rise of generative AI in 2023 further increased demand for GPU-based compute, with providers like AWS, Azure, and GCP integrating AI-specific chips and models such as OpenAI's solutions. Sustainability also emerged as a critical focus, with hyperscalers committing to renewable energy and carbon-neutral operations.

\subsection{Industrial-Scale Cloud Infrastructures}

The rise of cloud computing, driven by billions of smartphones converting into small computers that require storage and processing power from backend systems, has provided the elasticity and scalability necessary for modern AI applications. Hyperscale data centers, powered by services like AWS, Azure, and Google Cloud, have become the backbone of AI research and deployment.

The scale of these data center farms has grown significantly over the years. What once required energy measured in megawatts has now escalated to gigawatt levels, comparable to the energy generation of large nuclear power plants. This dramatic increase reflects the massive computational demands of modern AI workloads, including the training and deployment of large-scale models. Hyperscalers continue to invest heavily in optimizing energy efficiency while exploring sustainable energy sources to mitigate the environmental impact of such large-scale operations.

The topic of electricity consumption remains a focal point at energy conferences worldwide, often featuring packed sessions. Notably, while previous editions of the International Energy Agency's reports on electricity markets made no significant reference to data centers, this year's \textit{Electricity 2024: Analysis and Forecast to 2026} dedicates an entire section to their growing impact on global electricity consumption \cite{iea2024electricity}.

Cloud platforms have democratized access to computational resources, enabling small and large enterprises alike to leverage AI technologies. The integration of edge computing and serverless models further extends this scalability, bringing AI capabilities closer to users while optimizing resource use.

These innovations support not only AI research but also real-time applications, ensuring that powerful models can function effectively across diverse use cases, from conversational agents to autonomous systems.

\subsubsection{Implications for AI and Scalability}

Cloud services have become essential for AI, offering the scalability required for training and deploying large-scale models. By integrating advanced computing technologies, cloud providers enable businesses and researchers to access vast computational resources on demand. This synergy between cloud services and AI continues to drive innovation across industries, shaping the future of technology.

\subsection{Scientific Breakthroughs in AI}

\begin{figure}[!h]
    \centering
\includegraphics[width=0.5\textwidth]{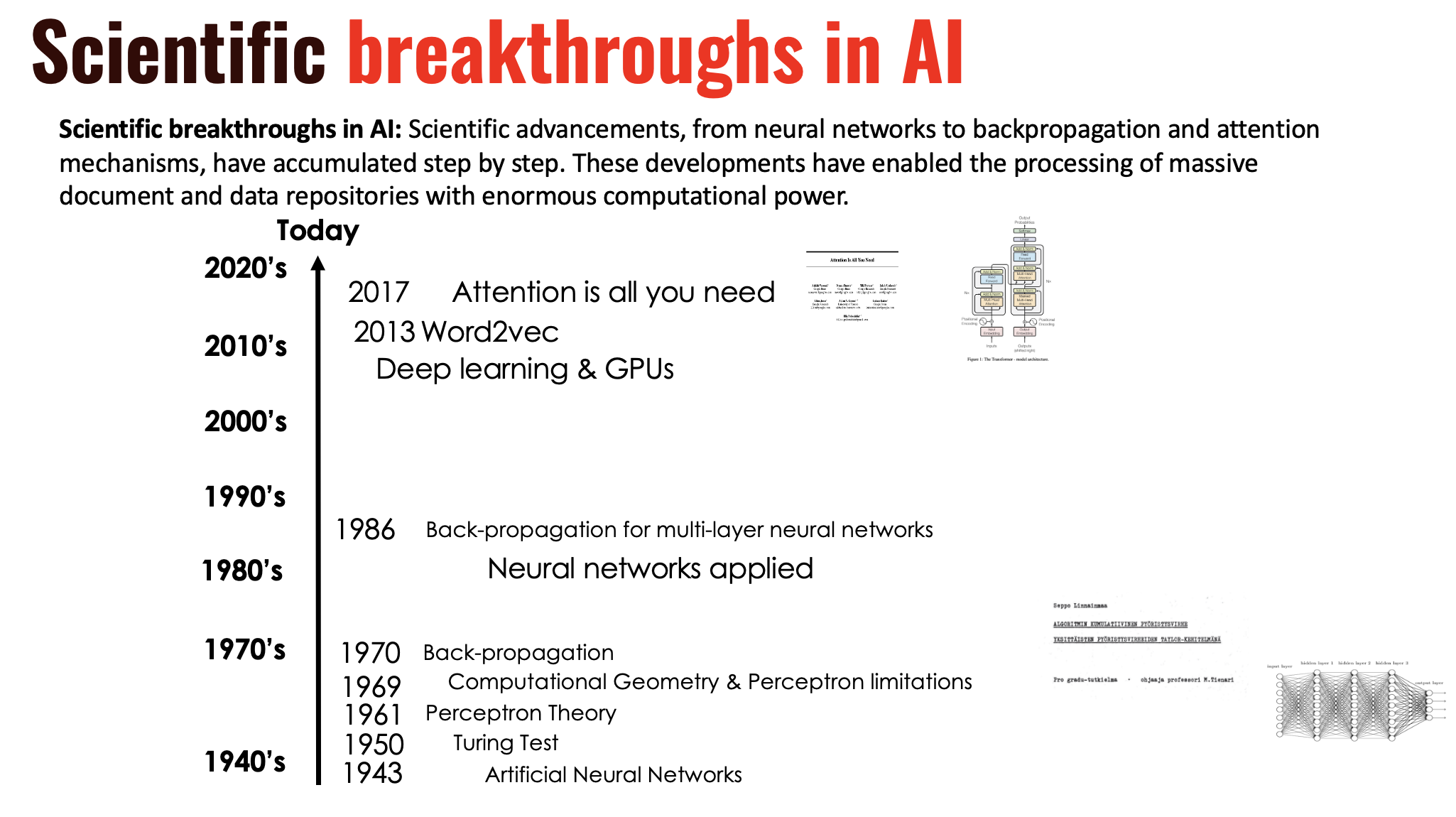}
	\caption{Scientific breakthroughs in AI: Scientific advancements, from neural networks to backpropagation and attention mechanisms, have accumulated step by step. These developments have enabled the processing of massive document and data repositories with enormous computational power.}
\label{fig:ai-research-history}
\end{figure}

Scientific advancements have paved the way for the remarkable progress seen in artificial intelligence (AI) today. From neural networks to backpropagation algorithms and transformer-based architectures, these breakthroughs have provided the foundation for processing massive data repositories with immense computational power.

\subsubsection{The Turing Test}

The Turing Test, introduced by Alan Turing in his 1950 paper \cite{turing1950computing}, was proposed as a way to assess whether a machine can exhibit behavior indistinguishable from that of a human. The test involves an interrogator communicating with both a human and a machine through a text-based interface, without knowing which is which. If the interrogator cannot consistently differentiate between the human and the machine, the machine is considered to have passed the test.

Turing's concept has sparked ongoing discussions about machine intelligence and remains a reference point in AI research. While modern AI systems often surpass the capabilities envisioned in Turing's era, the test continues to frame debates on the nature of intelligence, the goals of AI, and the ethical considerations surrounding the development of human-like machines.

\subsubsection{Neural Networks: A Digital Brain}

Neural networks, often described as "digital brains," are computational models inspired by the structure and function of biological neural systems. They consist of interconnected nodes or neurons organized into layers: input, hidden, and output. These layers work together to process and analyze data, enabling the network to identify patterns and make predictions.

The history of neural networks can be traced back to early computational models like the "McCulloch-Pitts neuron" \cite{mcculloch1943logical}, which introduced a formal framework for understanding neural computation. Building on this foundation, Rosenblatt \cite{rosenblatt1961principles} developed the perceptron model, a simple neural network framework inspired by biological neurons. The perceptron demonstrated the ability of single-layer networks to learn linearly separable patterns, establishing an early foundation for machine learning. However, Minsky and Papert \cite{minsky1969} later highlighted the limitations of perceptrons, particularly their inability to solve problems involving non-linear separability. This critique emphasized the need for more advanced architectures and methods.

These challenges were addressed in the 1980s with the introduction of backpropagation and the development of multilayer neural networks, which enabled the learning of complex, non-linear relationships. These innovations paved the way for modern deep learning, where neural networks with multiple hidden layers are used to model intricate patterns and solve a wide range of complex problems.

\textbf{Neural Network Key Features:}

\begin{itemize}
    \item \textbf{Learning Ability:} Neural networks are trained using data and improve their performance by iteratively adjusting their internal parameters.
    \item \textbf{Adaptability:} These models can be applied to a wide range of tasks, including image classification, speech recognition, and natural language processing.
    \item \textbf{Deep Learning:} By leveraging multiple hidden layers, deep learning models are capable of recognizing complex and high-dimensional patterns in data.
\end{itemize}

\subsubsection{Backpropagation Algorithm}

The backpropagation algorithm, first introduced in 1970 by Seppo Linnainmaa \cite{linnainmaa1970representation}, is a fundamental method for training neural networks. It uses the gradient of the error with respect to the network's weights to iteratively adjust the weights, thereby minimizing the error and improving the network's performance. The 1986 work by Rumelhart, Hinton, and Williams \cite{rumelhart1986learning} popularized the use of backpropagation for training multilayer networks, leading to renewed interest in neural networks during the 1980s and 1990s.

\textbf{Steps in Backpropagation:}
\begin{itemize}
    \item \textbf{Initialize Weights:} Begin by assigning random initial values to the network's weights.
    \item \textbf{Forward Pass:} Compute the output of the network for a given input by propagating the input through the layers.
    \item \textbf{Error Calculation:} Compare the predicted output with the actual output to calculate the error.
    \item \textbf{Backward Pass:} Propagate the error backward through the network to compute the gradient of the error with respect to each weight.
    \item \textbf{Weight Update:} Adjust the weights iteratively using gradient descent to minimize the error.
\end{itemize}

Backpropagation remains a critical algorithm for training deep neural networks, enabling advancements in fields such as computer vision, natural language processing, and speech recognition.

\subsubsection{The Role of Word2Vec in Training Transformer-Based LLMs}

Word2Vec, introduced by Mikolov et al. (2013)~\cite{mikolov2013efficient}, generated dense vector representations of words, capturing semantic and syntactic relationships. This marked a shift in natural language processing (NLP) from traditional sparse representations like one-hot encoding to distributed embeddings, enabling models to understand contextual meanings through vector proximity.

Though predating Transformer architectures, Word2Vec laid the foundation for embedding layers in models like GPT and BERT, where discrete tokens are converted into continuous representations for attention-based processing. Its skip-gram and continuous bag-of-words (CBOW) methods inspired later self-supervised learning approaches, such as masked language modeling and autoregressive training. 

Word2Vec's focus on scalability and distributed representations also paved the way for training on massive datasets, a cornerstone of modern LLMs. Its impact is evident in embedding-rich architectures that underpin state-of-the-art NLP systems.

\subsubsection{"Attention Is All You Need": The Transformer Architecture (2017)}

The 2017 paper \textit{Attention Is All You Need} introduced the Transformer model, fundamentally reshaping AI by replacing traditional recurrent and convolutional networks with a self-attention mechanism ~\cite{vaswani2017attention}. This architecture enabled parallel computation and scalable training on massive datasets.

\textbf{Key Contributions:}
\begin{itemize}
    \item \textbf{Scalability:} The highly parallelizable self-attention mechanism revolutionized natural language processing (NLP).
    \item \textbf{Foundation for LLMs:} Transformers are the backbone of advanced models like GPT, BERT, and T5, which excel in language understanding and generation.
    \item \textbf{General-Purpose Framework:} Beyond NLP, Transformers have influenced fields such as computer vision, protein folding, and reinforcement learning.
\end{itemize}

The Transformer architecture represents a turning point in AI, setting new standards for performance and scalability.

\subsubsection{Impact of Scientific Breakthroughs}

These breakthroughs, combined with advances in computational power and data availability, have enabled AI to process vast and complex datasets. They form the foundation of modern AI applications, empowering industries ranging from healthcare to autonomous vehicles. The integration of neural networks, backpropagation, and transformers continues to drive innovation, shaping the future of AI.

\subsubsection{Generative AI: Beyond the Transformer Architecture}
It is an oversimplification to attribute the rise and success of Generative AI (GenAI) solely to the limited set of named papers in this section. 

While the Transformer model is undoubtedly a critical component, it represents just one of many advancements that have collectively enabled the remarkable capabilities of GenAI. The development of Generative AI is a convergence of breakthroughs across numerous domains, including hardware scalability, large-scale dataset curation, algorithmic innovations, and the integration of multi-modal learning. Readers seeking a deeper understanding of the evolution and mechanics of Generative AI are encouraged to refer to the papers cited below.

For a comprehensive overview of the evolution, capabilities, and future prospects of large language models (LLMs) readers are advised to go through \cite{naveed2023comprehensive} and \cite{cao2023comprehensive}. 

The paper \cite{naveed2023comprehensive} discusses the role of large-scale pretraining, fine-tuning methodologies, and the challenges in scaling LLMs, highlighting the importance of reinforcement learning with human feedback (RLHF) and advancements in efficient training techniques. It emphasizes that the interplay between data size, model complexity, and computational resources is key to understanding the power and limitations of LLMs.

Additionally, the paper \cite{cao2023comprehensive} explores the broader history of GenAI, tracing its roots from early Generative Adversarial Networks (GANs) to the latest advancements in ChatGPT and related models. The survey identifies the progression of GenAI architectures and explains how advancements like self-attention, diffusion models, and multi-modal integration have expanded the scope of generative capabilities beyond text to include images, audio, and video.

\section{Discussion}

\subsection{Ambitious Research Objectives}
This study set forth an ambitious set of research questions and objectives, primarily aiming to identify the principal components and breakthroughs that have made the current age of AI applications possible. By tracing the historical and technological developments of AI, this research sought to provide a comprehensive understanding of how key innovations and milestones converged to enable the transformative capabilities of modern AI systems.

\subsection{Backward-Looking Focus}
It is important to emphasize that this paper adopts a backward-looking perspective, focusing exclusively on the historical and technical foundations of current AI advancements. It does not attempt to predict future developments in the field. While questions about the future trajectory of AI are undoubtedly compelling, they require a different research methodology, including scenario analysis, foresight studies, and trend forecasting. These topics are beyond the scope of this paper and could form the basis of a separate, forward-looking study.

\subsection{Limitations of Simplification}
Condensing the vast array of developments into a single paper inevitably involves harsh simplifications. The interplay of technological, scientific, and societal factors that shaped AI's trajectory is complex and multifaceted. While this paper highlights the most critical milestones, some nuances and interdependencies may not have been captured in sufficient detail. Future studies could aim to address these limitations by examining specific subdomains or advancements in isolation.

\subsection{Self-Critique and Missing Points}
Several areas of potential improvement and missing elements warrant reflection:
\begin{itemize}
    \item \textbf{Algorithmic Innovations:} While this paper highlights key breakthroughs such as neural networks, backpropagation, and the Transformer model, other important developments, such as advancements in optimization algorithms and reinforcement learning, were not comprehensively explored.
    \item \textbf{Ethical and Societal Implications:} The paper does not address the ethical, societal, and environmental challenges posed by the proliferation of AI technologies. These are critical areas for future research.
    \item \textbf{Diverse Data Ecosystems:} Although the role of the World Wide Web was discussed, the impact of proprietary datasets and domain-specific data sources remains underexplored.
    \item \textbf{Application-Specific Analysis:} This paper takes a generalist approach to AI applications. A more detailed exploration of verticals such as healthcare, autonomous systems, or education could provide valuable insights.
\end{itemize}

\subsection{Interdisciplinary Synergies}
AI's evolution has been profoundly influenced by interdisciplinary collaboration, combining insights from computer science, mathematics, engineering, and neuroscience. Exploring these synergies more deeply could uncover additional pathways for innovation and application.

\subsection{Broader Implications for Society}
The democratization of AI, as exemplified by applications accessible to children, underscores its transformative potential. However, it also raises critical questions about data privacy, the digital divide, and societal impact. These issues demand ongoing dialogue among researchers, policymakers, and the public.

\subsection{History and the Future}
This paper has provided a backward-looking analysis of the historical and technological factors that enabled the age of AI applications. While the focus was on tracing past developments, the question of AI's future trajectory remains an intriguing and separate area of study. Addressing it would require adopting different methodologies, such as trend forecasting and scenario analysis, and could form the subject of a dedicated paper. By shedding light on the foundational achievements of AI, this research lays the groundwork for both understanding the present and contemplating future possibilities.

\section{Conclusions}
\label{sec:conclusion}

This paper explored why we are currently living in the age of AI applications, situating the discussion within the context of technological evolution and a practical, real-world example of a child's interaction with an AI application. 

By addressing the research questions discussed in \ref{sec:research_questions} and the research objectives outlined in \ref{sec:research_objectives}, this study identifies five key factors as critical enablers for the emergence of these AI applications.

\subsection{Evolution of Computational Hardware}
The advent of advanced computational hardware, particularly CPUs and GPUs, has revolutionized the field of AI. These innovations have enabled the training of highly complex AI models, making previously unattainable tasks feasible. From the early development of microprocessors in the 1970s to the introduction of GPUs for general-purpose computing, each generation of hardware laid the groundwork for modern AI systems.

In particular, GPUs have emerged as indispensable for AI applications. Their ability to execute parallel computations has made them ideal for training deep neural networks. Recent advancements, such as NVIDIA's A100, have further optimized hardware for high-performance AI tasks, enabling breakthroughs in natural language processing and computer vision.

The integration of specialized accelerators, such as Google's Tensor Processing Units (TPUs), and the early explorations in quantum computing, highlight the continuous drive toward efficiency and scalability. These developments ensure that AI systems can handle increasing levels of complexity while remaining accessible to a broad audience.

\subsection{The Role of the World Wide Web (WWW) in Data Availability}
The World Wide Web has been a cornerstone for AI development, providing an unparalleled source of structured and unstructured data. From its inception in the 1990s to the exponential growth in data-driven content, the WWW has fostered an environment ripe for AI training and experimentation.

Search engines, particularly Google, transformed the way data was accessed and consumed, making the vast repositories of the web navigable and useful. The dynamic evolution from static web pages to rich, user-generated content further diversified the datasets available to AI researchers.

This surge in data availability has enabled the training of models capable of addressing complex tasks. It has also democratized AI, allowing applications like language models and recommendation systems to thrive in a variety of domains.

\subsection{Scale of Mobile Computing}
The transformation of mobile phones into powerful computational devices has significantly influenced the AI landscape. With over 6 billion smartphones globally, mobile devices serve as both data generators and interfaces for AI applications.

Advancements in mobile communication technologies, from 1G to 5G, have enhanced connectivity and data processing capabilities. The widespread adoption of Android and iOS platforms has created a seamless ecosystem for deploying AI-powered applications directly into users' hands.

The sheer scale of billions of smartphones acting as small computers has necessitated the development of enormous data storage and processing infrastructure. These hyperscale, industrial-scale cloud platforms have become the backbone of modern AI, enabling the collection, storage, and analysis of the vast amounts of data generated by mobile devices. This infrastructure ensures that AI systems can process and deliver insights in real time, meeting the demands of users worldwide.

This ubiquitous access to AI has not only driven innovation but also made sophisticated technologies accessible to even the youngest members of society. The story of a child creating bedtime stories using an AI app exemplifies this trend, showcasing how intuitive interfaces and advanced AI can transform everyday interactions.

\subsection{Industrial-Scale Cloud Infrastructures}
The rise of cloud computing, driven by billions of smartphones converting into small computers that require storage and processing power from backend systems, has provided the elasticity and scalability necessary for modern AI applications. Hyperscale data centers, powered by services like AWS, Azure, and Google Cloud, have become the backbone of AI research and deployment.

The scale of these data center farms has grown significantly over the years. What once required energy measured in megawatts has now escalated to gigawatt levels, comparable to the energy generation of large nuclear power plants. This dramatic increase reflects the massive computational demands of modern AI workloads, including the training and deployment of large-scale models. Hyperscalers continue to invest heavily in optimizing energy efficiency while exploring sustainable energy sources to mitigate the environmental impact of such large-scale operations.

The topic of electricity consumption remains a focal point at energy conferences worldwide, often featuring packed sessions. Notably, while previous editions of the International Energy Agency's reports on electricity markets made no significant reference to data centers, this year's \textit{Electricity 2024: Analysis and Forecast to 2026} dedicates an entire section to their growing impact on global electricity consumption.

Cloud platforms have democratized access to computational resources, enabling small and large enterprises alike to leverage AI technologies. The integration of edge computing and serverless models further extends this scalability, bringing AI capabilities closer to users while optimizing resource use.

These innovations support not only AI research but also real-time applications, ensuring that powerful models can function effectively across diverse use cases, from conversational agents to autonomous systems.

\subsection{Breakthroughs in AI Research}
Scientific breakthroughs have been fundamental to the success of AI applications. Key advancements, such as neural networks, backpropagation, and the Transformer architecture, have redefined the boundaries of what AI can achieve.

The development of the Transformer model, for example, introduced a highly parallelizable architecture that enabled large-scale language models like GPT. These innovations have transformed natural language processing, allowing AI to generate human-like text, translate languages, and understand context with remarkable accuracy.

While all other enablers, such as advancements in computational hardware, cloud infrastructures, and data availability, have been critical, the current level of sophistication seen in large language models (LLMs) would not have been possible without the Transformer breakthrough. This paper introduced the self-attention mechanism, which revolutionized AI by enabling scalability, efficiency, and a deeper understanding of complex relationships in data.

Pinpointing the most impactful papers in AI research, however, is not an easy task. The field has been shaped by thousands of foundational contributions spanning decades, each playing a crucial role in building the knowledge and tools we have today. While it would be possible to list many of these influential works, doing so would go against the research objectives of this study, which aim to focus on a minimal set of critical breakthroughs that have enabled the current state of AI applications.

The cumulative impact of these breakthroughs has made AI a versatile and transformative tool, capable of addressing challenges across industries. By building on these foundations, AI continues to evolve, promising new opportunities for innovation and societal benefit.

\section*{Final Thoughts}
The convergence of hardware advancements, data availability, mobile computing, cloud infrastructures, and scientific research has ushered in a new era of AI applications. The practical example of a child effortlessly creating bedtime stories underscores the transformative potential of these technologies, making complex systems accessible and impactful for all. As AI continues to shape our world, it is crucial to navigate this evolution with both ambition and responsibility.


\bibliographystyle{apalike}
{\small
\bibliography{references_short}}

\begin{thebibliography}{}

\bibitem[Berners-Lee, 1989]{berners_lee_1989}
Berners-Lee, T. (1989).
\newblock {Information Management: A Proposal}.
\newblock \\URL: http://www.w3.org/History/1989/proposal.html.

\bibitem[Brin and Page, 1998]{brin_page:1998}
Brin, S. and Page, L. (1998).
\newblock {The Anatomy of a Large-Scale Hypertextual Web Search Engine}.
\newblock {\em Computer Networks and ISDN Systems}, 30:107--117.
\newblock \\URL: http://www-db.stanford.edu/~backrub/google.html.

\bibitem[Bush, 1945]{vanner45}
Bush, V. (1945).
\newblock {\em As we may think}, pages 85--110.
\newblock ISBN: 0-12-523270-5, From Memex to hypertext: Vannevar Bush and the
  mind's machine. Academic Press Professional, Inc.
\newblock URL: http://www.theatlantic.com/ unbound/flashbks/computer/bushf.htm.

\bibitem[Cao et~al., 2023]{cao2023comprehensive}
Cao, Y., Li, S., Liu, Y., Yan, Z., Dai, Y., Yu, P.~S., and Sun, L. (2023).
\newblock A comprehensive survey of ai-generated content (aigc): A history of
  generative ai from gan to chatgpt.
\newblock {\em arXiv preprint arXiv:2303.04226}.

\bibitem[Engelbart and English, 1968]{engelbart:1968}
Engelbart, D.~C. and English, W.~K. (1968).
\newblock {\em {A Research Center For Augmenting Human Intellect}}, volume~33,
  pages 395--410.
\newblock Thompson Book Co, afips conference proceedings edition.

\bibitem[{IEA}, 2024]{iea2024electricity}
{IEA} (2024).
\newblock Electricity 2024: Analysis and forecast to 2026.

\bibitem[Linnainmaa, 1970]{linnainmaa1970representation}
Linnainmaa, S. (1970).
\newblock {\em The representation of the cumulative rounding error of an
  algorithm as a Taylor expansion of the local rounding errors}.
\newblock PhD thesis, Master’s Thesis (in Finnish), Univ. Helsinki.

\bibitem[McCulloch and Pitts, 1943]{mcculloch1943logical}
McCulloch, W.~S. and Pitts, W. (1943).
\newblock A logical calculus of the ideas immanent in nervous activity.
\newblock {\em The bulletin of mathematical biophysics}, 5:115--133.

\bibitem[Mikolov, 2013]{mikolov2013efficient}
Mikolov, T. (2013).
\newblock Efficient estimation of word representations in vector space.
\newblock {\em arXiv preprint arXiv:1301.3781}, 3781.

\bibitem[Minsky and Papert, 1969]{minsky1969}
Minsky, M. and Papert, S. (1969).
\newblock {\em Perceptrons: An Introduction to Computational Geometry}.
\newblock MIT Press, Cambridge, MA.

\bibitem[Naveed et~al., 2023]{naveed2023comprehensive}
Naveed, H., Khan, A.~U., Qiu, S., Saqib, M., Anwar, S., Usman, M., Akhtar, N.,
  Barnes, N., and Mian, A. (2023).
\newblock A comprehensive overview of large language models.
\newblock {\em arXiv preprint arXiv:2307.06435}.

\bibitem[Rosenblatt, 1961]{rosenblatt1961principles}
Rosenblatt, F. (1961).
\newblock Principles of neurodynamics: Perceptrons and the theory of brain
  mechanisms.

\bibitem[Rumelhart et~al., 1986]{rumelhart1986learning}
Rumelhart, D.~E., Hinton, G.~E., and Williams, R.~J. (1986).
\newblock Learning representations by back-propagating errors.
\newblock {\em nature}, 323(6088):533--536.

\bibitem[Turing, 1950]{turing1950computing}
Turing, A.~M. (1950).
\newblock Computing machinery and intelligence.
\newblock {\em Mind}, 59(236):433--460.

\bibitem[Vaswani et~al., 2017]{vaswani2017attention}
Vaswani, A., Shazeer, N., Parmar, N., Uszkoreit, J., Jones, L., Gomez, A.~N.,
  Kaiser, {\L}., and Polosukhin, I. (2017).
\newblock Attention is all you need.
\newblock {\em Advances in neural information processing systems}, 30.

\bibitem[{Wikipedia contributors}, 2024a]{wikipedia_android_history}
{Wikipedia contributors} (2024a).
\newblock Android version history.
\newblock \url{https://en.wikipedia.org/wiki/Android\_version\_history}.
\newblock Accessed: 2024-12-26.

\bibitem[{Wikipedia contributors}, 2024b]{wikipedia_gpu_history}
{Wikipedia contributors} (2024b).
\newblock Graphics processing unit.
\newblock \url{https://en.wikipedia.org/wiki/Graphics_processing_unit}.
\newblock Accessed: 2024-12-26.

\bibitem[{Wikipedia contributors}, 2024c]{wikipedia_history_linux}
{Wikipedia contributors} (2024c).
\newblock History of linux.
\newblock \url{https://en.wikipedia.org/wiki/History\_of\_Linux}.
\newblock Accessed: 2024-12-26.

\bibitem[{Wikipedia contributors}, 2024d]{wikipedia_mobile_os}
{Wikipedia contributors} (2024d).
\newblock Mobile operating system.
\newblock \url{https://en.wikipedia.org/wiki/Mobile\_operating\_system}.
\newblock Accessed: 2024-12-26.

\end{thebibliography}


\end{document}